\documentclass[prl,twocolumn,showpacs,%preprintnumbers,
superscriptaddress]{revtex4}
\usepackage[hypertex]{hyperref}
\usepackage{amsmath}
\usepackage{graphicx}
\usepackage{bm}
\usepackage{amsfonts}
\usepackage{amssymb}%

\def\Hn{\mathcal{H}_2^{\otimes n}}
\def\clg{\mathop{\mathrm{Cl}_{\mathcal{G}}}}
\def\Z{{\overline Z}}

\def\X{{\overline X}}
\def\Gn{\mathbb{P}_n}
\def\Wg{\mathbb{W}} % group for USt code
\def\span{\mathop{\rm span}}
\def\QA{{\mathcal{Q}_A}}% group of graph-images of correctable errors on A.
\def\DA{\mathbb{D}_A}% group of graph-images of correctable errors on A.
\def\EA{\mathbb{E}_A}% group of correctable errors on A.

\newcommand{\bra}[1]{\left\langle{#1}\right\vert}
\newcommand{\ket}[1]{\left\vert{#1}\right\rangle}

\newtheorem{theorem}{Theorem}

\begin{document}
%%% \advance\abovecaptionskip by -0.5em
%\advance\belowcaptionskip by -0.4em
%\advance\abovedisplayskip by -0.3em
%\advance\belowdisplayskip by -0.3em
%%% \advance\abovedisplayshortskip by -0.5em
%%% \advance\belowdisplayshortskip by -0.5em
%%% \advance\floatsep by -0.5em
%%% \advance\textfloatsep by -0.5em
%%% %\intextsep       14pt plus 4pt minus 4pt
%%% \advance\dblfloatsep by -0.5em
%%% \advance\dbltextfloatsep by -0.5em

\title{Clustered Error Correction of Codeword-Stabilized Quantum Codes}
\date{\today} 
\author{Yunfan Li} 
\affiliation{Department of Electrical Engineering, University of
  California, Riverside, California 92521, USA} 

\author{Ilya Dumer} 
\affiliation{Department of Electrical Engineering, University of
  California, Riverside, California 92521, USA} 

\author{Leonid P. Pryadko} 
\affiliation{Department of Physics \& Astronomy, University of
  California, Riverside, California 92521, USA}

\begin{abstract}
  Codeword stabilized (CWS) codes are a general class of quantum codes
  that includes stabilizer codes and many families of non-additive
  codes with good parameters.  For such a non-additive code correcting
  all $t$-qubit errors, we propose an algorithm that employs a single
  measurement to test all errors located on a given set of $t$
  qubits.  Compared with exhaustive error screening, this reduces the
  total number of measurements required for error recovery by a factor
  of about $3^t$.
\end{abstract}

 \maketitle

%% 03.67.Pp Quantum error correction and other methods for protection
%% against decoherence
%% 03.67.Ac Quantum algorithms, protocols, and simulations
%% 89.70.Kn Channel capacity and error-correcting codes

 Quantum computation admits polynomial
complexity for many classical algorithms believed to be hard
\cite{Shor-factoring-1994,Nielsen-book}. To preserve coherence,
quantum computations must be protected by quantum error correcting
codes \cite{shor-error-correct,Knill-Laflamme-1997}.  Stabilizer codes
\cite{gc} represent an important class of quantum codes that 
can be encoded and decoded  in  polynomial time.
Recent Refs.~\onlinecite{Smolin-2007,Cross-CWS-2009} 
introduce a 
larger class of codeword-stabilized (CWS) codes.  It includes
important code families, such as the stabilizer codes and generally
non-additive union stabilizer (USt)
codes~\cite{ust}.
%\cite{Grassl-1997,Grassl-Roetteler-2008A,Grassl-Roetteler-2008B}.
CWS codes have a broader range of code parameters which can be
superior to those of any stabilizer
code\cite{Smolin-2007,Cross-CWS-2009,Chen-Zeng-Chuang-2008,ust}.
%\cite{,% 
%  Grassl-Roetteler-2008A,Grassl-Roetteler-2008B,Grassl-2009}.  

The most important advantage of the CWS codes is their close relation with the
classical codes.  In particular, a qubit CWS code $\mathcal{Q}$ can be mapped
onto a classical binary code $\mathcal{C}$, with the quantum Pauli errors also
mapped into some binary error patterns \cite{Cross-CWS-2009}.  This
way, within CWS framework, 
quantum code design can be reduced to classical codes and employ the wealth of
different techniques developed for the latter.

On the other hand, quantum error correction must preserve the original
quantum state in all intermediate measurements, and therefore is more
restrictive than many classical algorithms.  Thus, design of CWS codes
must be complemented by an efficient non-damaging quantum error
correction algorithm.  In this paper, our main goal is to address this
important unresolved problem.

We consider a general non-additive CWS code $((n,K,d))$ of distance $d$ which
encodes $K$ quantum dimensions into a $K$-dimensional subspace of the Hilbert
space of $n$ qubits.  This code detects all errors that corrupt up to $(d-1)$
qubits, and corrects all errors corrupting
$t\equiv\left\lfloor (d-1)/2\right\rfloor$ or fewer qubits. As a benchmark for
our study, we consider generic algorithms that project a corrupted code state
into different subspaces.  This brute-force technique is similar to the
exhaustive error screening in nonlinear classical codes, and requires up to%
\begin{equation}
B(n,t)=\sum_{i=0}^{t} \binom{n}{i}\,3^{i}\label{sphere}%
\end{equation}
measurements to screen all errors of weight $t$ or less.

To reduce the number of such measurements, we first design an error
detection algorithm for USt
codes\cite{ust}.
%%\cite{Grassl-1997,Grassl-Roetteler-2008A,Grassl-Roetteler-2008B}.
In CWS framework, the  classical code $\mathcal{C}$
associated with the USt code $\mathcal{Q}=((n,m\,2^k,d))$ is decomposed as a
group $\mathcal{C}_0$ of $2^{k}$ codewords shifted by $m$ binary
``translation'' vectors.  We prove the following
\begin{theorem}
  \label{complexity} For a USt code of length $n$, with a group of size
  $2^{k}$ and dimension $K=m\,2^{k}$, an error-detecting measurement requires
  no more than $2m(n-k)(n+3)$ two-qubit gates.
\end{theorem}

Then, for a general CWS code $\mathcal{Q}$, we propose an
error-correcting method that simultaneously screens all $4^{t}$
different errors located on any given subset of $t$ qubits, by
designing an auxiliary USt code which uses binary maps of these errors
as generators of the group $\mathcal{C}_0$, and the codewords of the
associated classical code $\mathcal{C}$ as translations.  This
requires only ${\binom{n}{t}-1}$ measurements to screen all groups.
Once the corrupted qubits are located, we need up to $2t$
extra measurements to find the actual error within the group.
Overall, this reduces the number $B(n,t)$ of measurements about
$3^{t}$ times to
\begin{equation}
N(n,t)={\binom{n}{t}}+2t-1. \label{eq:measur-new}%
\end{equation}
Our main result is summarized as

\begin{theorem}
\label{main}Consider any $t$-error correcting  CWS
code of length $n$ and dimension $K$.  Then this code can correct
errors using at most $N(n,t)$ measurements, each of which requires at
most $2K(n-1)(n+3)$ two-qubit gates.
\end{theorem}

\textsc{Definitions.} Throughout the paper, we use the Hilbert space
$\Hn\equiv {\mathcal{H}}^{2^{n}}$ to represent any $n$-qubit state.  Also,
$\Gn=\pm\{ 1, i\}\,\{I,X,Y,Z\}^{\otimes n}$ denotes the Pauli group,
where the number of non-trivial terms in the tensor product is
the {\em weight\/}  of a given $E\in\Gn$. 
We say that a
space $\mathcal{P}$ is \textit{stabilized} by a {\em measurement\/} operator
$M$ with all eigenvalues $\lambda=\pm1$ (this includes all Hermitian operators
in $\Gn$) if $M\ket\psi=\ket\psi$ for any state $\ket\psi$ in $\mathcal{P}$.
We will also use the term {\em anti-stabilized\/} if $M\ket\psi=-\ket\psi$.  A
space is stabilized by a set $\mathcal{M}$ of measurement operators if it is
simultaneously stabilized by all operators in $\mathcal{M}$. A maximal space
stabilized by $\mathcal{M}$ is called the \textit{stabilized} space
$\mathcal{P} (\mathcal{M})$, and $\mathcal{M}$ is called a stabilizer of
$\mathcal{P} (\mathcal{M})$.  The corresponding projector is denoted
$P_\mathcal{M}$.  The projector $\openone-P_\mathcal{M}$ corresponds to the
orthogonal complement $\mathcal{P}^\perp(\mathcal{M})$.  For a single
measurement operator, $M=2P_M-\openone$.

A general {\em quantum code\/} $((n,K,d))$ is a subspace $\mathcal{Q}\in \Hn$
of dimension $K$, such that any {\em detectable\/} error either takes any
non-zero state $\ket\psi\in\mathcal{Q}$ into a state outside of
$\mathcal{Q}$, 
$E\ket\psi\not\in\mathcal{Q}$, or acts trivially on
$\mathcal{Q}$,   %for any $\ket\psi\in \mathcal{Q}$,
$E\ket\psi=C_E\ket\psi$ with $C_E$ independent of $\ket\psi$.  A
combination $E_1^\dagger E_2$ of any two errors from a set
$\mathcal{E}$ of {\em correctable\/} errors is detectable.  The errors
are in the same \emph{degeneracy class} iff $E_1^\dagger E_2$ acts
trivially on $\mathcal{Q}$.  For a distance-$d$ code, all Pauli errors
of weight up to $(d-1)$ are detectable, and all Pauli errors of weight
up to $t=\lfloor (d-1)/2\rfloor$
are correctable\cite{Knill-Laflamme-1997,Nielsen-book}.

A {\em stabilizer code\/}\cite{gc} %\cite{gottesman-thesis,Calderbank-1997}
$[[n,k,d]]$ is defined as the stabilized space of an Abelian group
$\mathbb{S}\equiv \langle G_1,\ldots, G_{n-k}\rangle$ of size
$2^{n-k}$, $-\openone\not\in\mathbb{S}$,  generated by
Hermitian Pauli operators $G_i$, $i=1,\ldots n-k$.  Explicitly,
\begin{equation}
  \mathcal{Q}=\{\ket\psi:S\ket\psi=\ket\psi,\ \forall\ 
  S\in\mathbb{S}\}.\label{eq:def-stabilizer-code}
\end{equation}
The {\em logical\/} operators $\X_i$, $\Z_i$, $i=1,\ldots, k$ commute
with the code stabilizer $\mathbb{S}$; they obey the usual Pauli
commutation relations.  These operators, along with $G_i\in\mathbb{S}$
and the trivial $i\openone$, serve as generators of the code {\em
  normalizer\/} $\mathbb{N}$, a group of operators $U\in\Gn$ that
preserve the stabilizer $\mathbb{S}$ under conjugation, $U S U^\dagger
=S$, $U\in \mathbb{N}$, $S\in\mathbb{S}$.  Each correctable error
$E\in\Gn$ acting non-trivially on the code anti-commutes with at least
one generator $G_{i}$, and correctable errors in different degeneracy
classes anti-commute with different subsets of $\mathbb{S}$.  The
corrupted code $E(\mathcal{Q})\equiv
\{E|\psi\rangle:|\psi\rangle\in\mathcal{Q}\}$ is anti-stabilized by
those generators $G_{i}$ that anti-commute with $E$.  Thus, a
stabilizer code can be corrected by measuring the generators $G_{i}$;
the corresponding set of eigenvalues $\lambda_i=\pm1$ forms the {\em
  syndrome\/} of the error.

A \textit{codeword-stabilized} (CWS)
code\cite{Smolin-2007,Cross-CWS-2009} $((n,K,d))$ 
is defined in terms of a {\em stabilizer state\/} $\ket s$ (which is an
$[[n,0]]$ stabilizer code), and a set of $K$ mutually commuting codeword
operators $\mathcal{W}\equiv\{W_{i}\}_{i=1}^{K} \subset\Gn$.  
Explicitly,% %the CWS code%
\begin{equation}
  \mathcal{Q}=\span(\{|w_{i}\rangle\}_{i=1}^{K},\;|w_{i}\rangle
  \equiv W_{i}|s\rangle).\label{eq:def-cws-code}
\end{equation}
The stabilizer
$\mathbb{S}\equiv\langle S_1,\ldots, S_n\rangle$ of the state $\ket s$
is the maximal Abelian subgroup of the Pauli group such that
$-\openone\not\in\mathbb{S}$; in the context of CWS codes it is called {\em
  word stabilizer\/}\cite{Cross-CWS-2009}.  

A CWS code is a stabilizer code iff the $K$ word operators $W_{i}$
form an (Abelian) group\cite{Cross-CWS-2009}.  Such a CWS code is called
additive; in this case $K=2^k$ with integer $k$.

A {\em union stabilizer\/} (USt)
code\cite{ust}
%%\cite{Grassl-1997,Grassl-Roetteler-2008A,Grassl-Roetteler-2008B}
can be defined as a CWS code $\mathcal{Q}$ whose word operators
contain a group,
\begin{equation}
  \mathcal{W}=\{t_j\prod_{i=1}^{k}g_i^{\alpha_{i}}:
  j=1,\ldots,m,\;\alpha_{i}\in\{0,1\}\}.
  \label{eq:def-ust}
\end{equation}
Here $g_i$ are generators of the group $\Wg\equiv\langle
g_{1},\ldots, g_{k}\rangle$ forming an additive code
$\mathcal{Q}_{0}=\mathop{\rm span}\left( \{W|s\rangle
  \}_{W\in\Wg}\right)$ with dimension $K_0=2^k$.  The
operators $t_j$ form a set $\mathcal{T}$ of $m$ {\em
  translations\/} for the code $\mathcal{Q}_0$.  The translated spaces
$t_j(\mathcal{Q}_0)$ are mutually orthogonal, which implies that  
the dimension of the code $\mathcal{Q}$ is $K=m\,2^k$.

The {\em standard form\/} of a CWS
code\cite{Smolin-2007,Cross-CWS-2009} is defined in terms of a graph
$\mathcal{G}$ with $n$ vertices and a classical code $\mathcal{C}$
containing $K$ binary codewords ${\bf c}_i$ of length $n$.  The graph
adjacency matrix $R\in \{0,1\}^{n\times n}$ defines the generators of
the stabilizer, $S_i \equiv X_i Z_1^{R_{i1}} Z_2^{R_{i2}}\ldots
Z_n^{R_{in}}$, while the classical codewords define the codeword
operators $W_i= Z^{{\bf c}_i}\equiv Z_1^{c_{i1}} \ldots Z_n^{c_{in}}$.
Most importantly, the graph relates the error-correction
properties\cite{Cross-CWS-2009} of the quantum CWS code
$\mathcal{Q}=(\mathcal{G},\mathcal{C})$ and the classical code
$\mathcal{C}$.  Indeed, the action of a single-qubit error $X_i$ on
the code is equivalent (up to an overall phase) to that of $X_i
S_i=Z_1^{R_{i1}}\ldots Z_n^{R_{in}}$.  Any Pauli operator $E=Z^{\bf
  v}X^{\bf u}$ can thus be mapped (up to a phase) to the operator
$Z^{\clg(E)}$.  Here the function
\begin{equation}
  \clg(E)\equiv\mathbf{v}+
  u_1 {\bf R}_1+ u_2 {\bf R}_2+ \ldots + u_n {\bf R}_n
  \;({\rm mod}\; 2)
  \label{binary}
\end{equation}
defines the {\em graph-induced\/} classical (binary) map  of $E$.
This function also defines the error degeneracy classes of the code
$\mathcal{Q}$: two correctable quantum errors $E_1$ and $E_2$ are
mutually degenerate iff $\clg(E_1)=\clg(E_2)$
\cite{Cross-CWS-2009,Chuang-CWS-2009}.  For a pair of correctable
errors $E_1$, $E_2$ from different degeneracy classes,
$\clg(E_1)\neq\clg(E_2)$, the corrupted spaces are always orthogonal, 
$E_1(\mathcal{Q})\perp
E_2(\mathcal{Q})$ \cite{Li-Dumer-Pryadko-long-2009}.

Any CWS code is locally Clifford-equivalent to a code in standard
form\cite{Cross-CWS-2009}.  For CWS codes in standard form, we will
denote the corresponding set of word operators and word stabilizer as
$\mathcal{W}_\mathcal{G}$ and $\mathbb{S}_\mathcal{G}$, respectively.

\textsc{Exhaustive screening for CWS codes.} 
We can detect errors  by measuring the operator 
$M_Q\equiv 2P_Q-\openone$,% 
\begin{equation}
  P_\mathcal{Q}\equiv \sum_{W\in\mathcal{W}}
  W\ket s\bra sW^{\dag}.
  \label{eq:cws-projector}
\end{equation}
The corresponding ancilla
measurement circuit which uses $2K[n^2+\mathcal{O}(n)]$ two-qubit
gates can be constructed as the special case of
Eq.~(\ref{finaldecomposition}) below.  A different circuit which
requires up to $n^2+K\mathcal{O}(n)$ two-qubit gates is constructed in
Ref.~\onlinecite{Li-Dumer-Pryadko-long-2009}.

The operators $EM_\mathcal{Q} E^\dagger$ stabilize the spaces
$E(\mathcal{Q})$.  For a CWS code $\mathcal{Q}$, these spaces are
orthogonal for mutually non-degenerate correctable errors $E$.  This
implies that an error can be located by measuring such operators for
$E$ from different degeneracy classes.
For a $t$-error correcting
code we can exhaustively test all correctable errors using up to
$B(n,t)$ measurements [Eq.~(\ref{sphere})].  This bound is tight for
{\em non-degenerate\/} codes where all linearly-independent
correctable errors are mutually non-degenerate.

\begin{figure*}[htbp]%
  \includegraphics[scale=1.0]{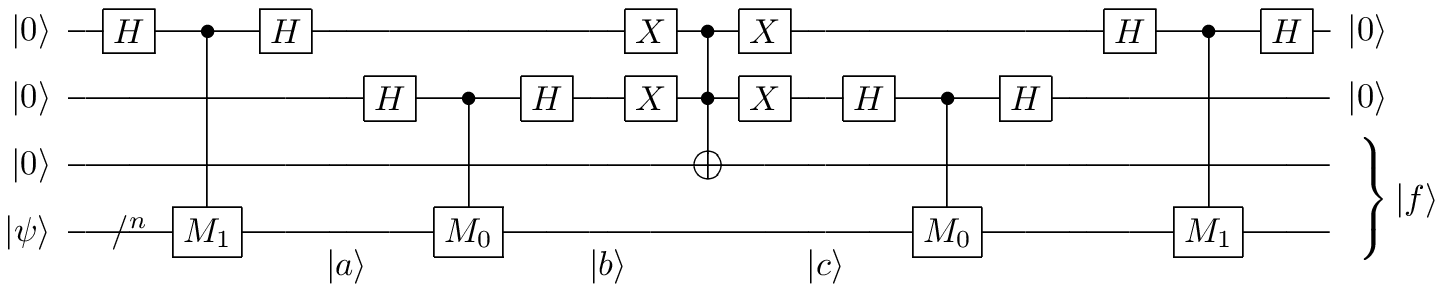}
  \caption{Measurement
    $M_{1}\wedge M_{0}$.  We use the
    projectors $P_i\equiv P_{M_i}$, $Q_i\equiv \openone-P_{i}$, $i=0,1$, and
    assume $P_1  P_0=P_0P_1$.  Circuit returns $\ket f=\ket{1} P_{1} P_{0}
    \ket\psi+\ket0(\openone-P_{1} P_{0})\ket\psi$, which is equivalent
    to $\ket f=\ket 1 P_{M_1\wedge M_0}+\ket 0(\openone-P_{M_1\wedge M_0})$. 
    Intermediate results: $\ket
    a=\ket{000}P_{1}\ket\psi+\ket{100}Q_{1}\ket\psi$, 
    $\ket b=\ket{000} P_{1} P_{0}\ket\psi+\ket{010}P_{1} Q_{0}\ket\psi+
    \ket{100}Q_{1} 
    P_{0}\ket\psi+ \ket{110}Q_{1} Q_{0}\ket\psi$, $\ket c=\ket{001} P_{1}
    P_{0}\ket\psi+\ket{010}P_{1} Q_{0}\ket\psi+ \ket{100}Q_{1} P_{0}\ket\psi+ \ket{110}Q_{1}
    Q_{0}\ket\psi$; the last two gate groups disentangle the first two
    ancillas.} 
\label{fig:measurementAnd}%
\end{figure*}

\begin{figure}[htbp]
  \centering \includegraphics[scale=1.0]{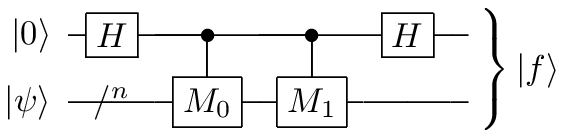}
  \caption{Measurement for $M_1\oplus M_0$.  Notations as in
    Fig.~\ref{fig:measurementAnd}. The result $\ket
    f=\ket1({Q}_1{P}_0+{P}_1{Q}_0) \ket\psi
    +\ket0({P}_1{P}_0+{Q}_1{Q}_0)\ket\psi$ is equivalent to
    $\ket1{P}_{M_1\oplus M_0}\ket\psi+\ket0(\openone-{P}_{M_1\oplus
      M_0})\ket\psi$.}
  \label{fig:measurementXor}
\end{figure}

\textsc{Measurement algebra}.  To simplify error correction, we will
first decompose multi-qubit measurements using the algebra of
projection operators \cite{Aggarwal-Calderbank-2008,Chuang-CWS-2009}.
A measurement $M$ projects a state into the stabilized space
$\mathcal{P}(M)$ or its orthogonal complement.  In the following we
assume that all measurement operators commute.   

In analogy with logical AND, let $M_{1}\wedge M_{0}$ denote the
measurement that stabilizes
$\mathcal{P}(M_{1})\cap\mathcal{P}(M_{0})$.  The circuit in
Fig.~\ref{fig:measurementAnd} shows an implementation of this
combination using logical operations on ancillas.
A different
circuit which requires only two ancillas is given in
Ref.~\onlinecite{Li-Dumer-Pryadko-long-2009}.  

An operation analogous to logical XOR is defined in terms of the
\emph{symmetric difference} of vector spaces $A\bigtriangleup B
\equiv \span(A\cap B^\perp,B\cap A^\perp)$.
We assume that there exists
an orthogonal basis common to all spaces.  Then, the symmetric difference
$A\bigtriangleup B\bigtriangleup C\bigtriangleup \ldots$ is spanned by the
basis vectors which belong to an odd number of subspaces $A, B, C, \ldots$.
We define the XOR of two commuting measurements, $M_{1}\oplus M_{0}$, as the
measurement that stabilizes
$\mathcal{P}(M_{1})\bigtriangleup\mathcal{P}(M_{0})$.  The corresponding
circuit [Fig.~\ref{fig:measurementXor}] is based on the easy-to-check identity
$M_{1}\oplus M_{0}=-M_{1} M_{0}$.

Generally, the equality symbol will denote the equivalence between
measurements.  If $M_1M_0=M_0M_1$, then%
\begin{eqnarray}
  \label{eq:m_and}
  M=M_1\wedge M_0&\Leftrightarrow& \mathcal{P}(M)=\mathcal{P}(M_1)\cap
  \mathcal{P}(M_0) ,\\
  \label{eq:m_xor}
  M=M_1\oplus M_0&\Leftrightarrow& \mathcal{P}(M)=\mathcal{P}(M_1)\bigtriangleup
  \mathcal{P}(M_0) .
\end{eqnarray}

\textsc{Decomposition of an additive CWS code}.  Consider an additive CWS
code $\mathcal{Q}_0$ with the set of word operators
$\mathcal{W}_0\equiv \mathbb{W}_0=\langle g_1,\ldots,g_k\rangle$
forming a group.  This code is a stabilizer code\cite{Cross-CWS-2009};
it is the common stabilized space of the $n-k$ generators $G_i$ of the
code stabilizer $\mathbb{S}_0$,
$\mathcal{Q}_0=\bigcap_{i=1}^{n-k}{\mathcal{P}(G_i)}$.  According to
Eq.~(\ref{eq:m_and}), we also have
\begin{equation}
  M_0\equiv M_{\mathcal{Q}_0}=\bigwedge_{i=1}^{n-k} {G_i},
  \label{eq:additive-decomposition}
\end{equation}
and can construct the corresponding measurement circuit by analogy
with Fig.~\ref{fig:measurementAnd} using associativity.
%%%, $M_2\oplus M_1\oplus M_0=M_2\oplus(M_1\oplus M_0)$.  
This requires
$2(n-k)$ controlled $n$-qubit Pauli operators and $(n-k-1)$
three-qubit Toffoli gates.  Adding the corresponding complexities
\cite{Bennett-1996}, we obtain the overall complexity of up to $2(n-k)
(n+3)$ two-qubit gates.

This measurement can be done in the basis of the original CWS
code.  The $n$ generators $S_{i}\in\Gn$ of the word stabilizer
$\mathbb{S}$ can be chosen \cite{Li-Dumer-Pryadko-long-2009} to
satisfy the {\em orthogonality\/} condition $S_{i}
g_{j}=(-1)^{\delta_{ij}}g_{j}S_{i}$.  Now, the $k$ logical operators
of the code can be chosen as $\X_j=g_j$, $\Z_j=S_j$, and the remaining
generators of the orthogonalized word stabilizer can serve as the
generators $G_i=S_{i+k}$, $i=1,\ldots, n-k$ of the code stabilizer
$\mathbb{S}_0$.

\textsc{Decomposition of a {\rm USt} code}.  Now consider a USt code
$\mathcal{Q}$ with the set $\mathcal{W}$ of word operators
in the form (\ref{eq:def-ust}).  Given the generators $G_i$ of the
stabilizer $\mathbb{S}_0$ of the additive subcode  $\mathcal{Q}_0$, the
generators of the translated code $t_j(\mathcal{Q}_0)$ can be
written as 
$t_j G_i t_j^\dagger$.  Then, the corresponding measurement operators
[cf.\ Eq.~(\ref{eq:additive-decomposition})] 
\begin{equation}
 M_j\equiv t_j M_{0}t_j^\dag=\bigwedge_{i=1}^{n-k}t_j G_i t_j^\dagger.
\end{equation}
The code $\mathcal{Q}$ is spanned by the orthogonal vector spaces% 
\begin{equation} 
  \mathcal{Q}\equiv \mathcal{P}(M_\mathcal{Q})
  =\span\left\{\mathcal{P}(M_{j})\right\}_{j=1}^{m},\;\,
  \mathcal{P}(M_{i})\perp\mathcal{P}(M_{j}), 
\end{equation}
which is equivalent to the symmetric difference
$\mathcal{Q}=\mathcal{P}(M_1)\bigtriangleup\mathcal{P}(M_2)
\bigtriangleup \ldots \bigtriangleup\mathcal{P}(M_{m})$.  According to
Eq.~(\ref{eq:m_xor}), this is also equivalent to the decomposition
\begin{equation}
  M_\mathcal{Q}=\bigoplus_{j=1}^{m}M_{j}=
  \bigoplus_{j=1}^{m}\left[  {\bigwedge_{i=1}^{n-k}}\left(  
        t_j G_i t_j^\dag\right)  \right]. 
  \label{finaldecomposition}
\end{equation}

Since the XOR (``$\oplus $'') of several measurements is implemented
as concatenation [Fig.~\ref{fig:measurementXor}], it requires no
overhead; the resulting complexity is then given by Theorem
\ref{complexity}.

\textsc{Clustered measurements for CWS codes}.  For a $t$-error correcting CWS
code $\mathcal{Q}$, consider any subset of correctable errors,
$\mathcal{E}'\subset\mathcal{E}$, and any correctable error $E$ not degenerate
with those in $\mathcal{E}'$.  Then, the space
$\mathcal{E}'(\mathcal{Q})\equiv \span_{E\in\mathcal{E}'}E(\mathcal{Q})$ is
orthogonal to $E(\mathcal{Q})$.  Furthermore, errors located on any $t$ qubits
(specified by the set of qubit indices $A=\{i_{1} ,\ldots i_{t}\}$) form a
group of correctable errors $\EA\equiv\langle X_{i},Z_{i}\rangle_{i\in A}$.
Thanks to the group property of $\EA$,
for the set $\mathcal{E}'\equiv \EA$, we also
have \cite{Li-Dumer-Pryadko-long-2009} a more restrictive 
identity $E(\QA)\perp\mathcal{Q}_{A}$, where $\QA\equiv \EA(\mathcal{Q})$.
Thus, $\QA$ is a quantum code which can detect errors $E\in\mathcal{E}$ not
degenerate with those in $\EA$.

Our clustered measurement technique is based on the observation that
$\QA$ is actually a USt code.  Indeed, consider the original
CWS code in standard form, $\mathcal{Q}=(\mathcal{G}, \mathcal{C})$.
The set of operators $\mathbb{D}_A\equiv \{Z^{\clg(E)}:E\in
\mathbb{E}_A\}$ forms an Abelian group of size $2^k\le |\EA|=2^{2t}$
since the operators $Z^{\clg(E)}$ obey the same multiplication table
as $E\in\EA$ but are not necessarily independent.  By construction,
different elements of $\DA$ are in different error degeneracy classes,
therefore the spaces $e_i(\mathcal{Q})$ are mutually orthogonal for
different $e_i\in\DA$.  The additional degenerate elements in $\EA$ do
not add to the span, therefore $\QA\equiv
\EA(\mathcal{Q})=\DA(\mathcal{Q})$.  Since $\DA$ and
$\mathcal{W}_\mathcal{G}$ are combinations of $Z$-operators only,
$\QA$ is a USt code in standard form which uses the same
stabilizer state $\ket s$ as $\mathcal{Q}$, the Abelian group
$\Wg=\DA$, and the codeword operators $\mathcal{W}_\mathcal{G}$ of the
code $\mathcal{Q}$ as the translation set $\mathcal{T}$
[Eq.~(\ref{eq:def-ust})].

To form the measurement $M_A\equiv M_\QA$ that stabilizes the USt code $\QA$, we
construct a set of $(n-k)$ orthogonal generators $G_i$ for the
additive code $\mathcal{Q}_0\equiv\DA(\span{\ket s})$, see
Eq.~(\ref{eq:additive-decomposition}).  The actual measurement [cf.\ 
  Eq.~(\ref{finaldecomposition})],
\begin{equation}
M_{A}=\bigoplus_{W\in{\mathcal{W}_\mathcal{G}}}{\left[  \bigwedge_{i=1}^{n-k
}{\left(  WG_{i}W^{\dag}\right)  }\right]  },\label{decomposeMA}%
\end{equation}
satisfies the complexity bound of Theorem~\ref{complexity}.  The
measurement $M_A$ has eigenvalue $1$  for all states in
$\QA$, and $-1$  for all states in
$\QA^\perp$, which corresponds to all correctable errors not
degenerate with those in $\EA$.  

To determine the error, we first perform measurements $M_{A}^{(j)}$ for
all (but the last one) size-$t$ index sets $A^{(j)}$. After locating the
covering cluster $A$ with Abelian group $\DA$ of size $|\DA|=2^s\le
2^{2t}$, we can find the error by going over all $s\le 2t$ subgroups
of $\EA$ with $s-1$ generators.  Each measurement determines whether or not 
the omitted generator is a part of the error.  The error is 
identified as a product of the generators present in all 
auxiliary codes that
detected no errors.  Overall, this requires up to $N(n,t)$
measurements as in Eq.~(\ref{eq:measur-new}).  Thus, for any code
length $n\geq3$, the former number of $B(n,t)$ measurements [see
  Eq.~(\ref{sphere})] is reduced by a factor
\begin{equation}
B(n,t)/N(n,t)\geq\left\{
\begin{array}
[c]{ll}%
\frac{3n+1}{n+1}, & \text{if}\;t=1,\smallskip\smallskip\\
\;3^{t}, & \text{if}\;t>1.
\end{array}
\right.  \label{factoreven}%
\end{equation}

 Some additional acceleration can be gained if the
original CWS code is a USt code, with the set of codeword
operators~(\ref{eq:def-ust}).  In this case, for a given index set
$A$, our scheme employs a bigger group ${\Wg}'$ which 
includes the
generators of both $\DA$ and the original group $\Wg$, and
a smaller translation set $\mathcal{T}$ of size $m<K$.  The 
complexity of a single measurement would then be reduced to $2mn^2$, compared
to $2K n^2$ in Theorem~\ref{main}.  Screening of $N(n,t)$ or fewer qubit
clusters will locate the error.

Note also that in the special case of stabilizer codes, our error-grouping
technique is equivalent to the syndrome-based
recovery\cite{Li-Dumer-Pryadko-long-2009}.  Indeed, for a stabilizer
code $\mathcal{Q}=[[n,k,d]]$, the degeneracy classes form  an Abelian
group $\mathbb{E} =\langle 
e_1,\ldots, e_{n-k}\rangle$ whose $2^{n-k}$ elements are enumerated by
different syndromes \endnote{This group is the quotient group of the
  Abelian version of $\Gn$ which ignores the phases, by the Abelian
  version of the code normalizer $\mathbb{N}(\mathcal{Q})$.}.
To locate the error, we can go over all $(n-k)$ USt codes
$\mathbb{E}_\alpha(\mathcal{Q})$ generated by the subgroups of
$\mathbb{E}$ with one generator, $e_\alpha$, missing.  Then, the code
$\mathbb{E}_\alpha(\mathcal{Q})$ is a stabilizer code that has to
correct only one non-trivial error, $\mathcal{E}_\alpha=\langle
e_\alpha \rangle$.  The corresponding stabilizer $\mathbb{S}_\alpha$
has only one generator.  Thus, error can be located by independent
measurements of $n-k$ Pauli operators, as we do to measure the
syndrome.

In conclusion, we constructed an accelerated {\em clustered\/} quantum
error correction algorithm for a non-additive CWS code which uses a set
of auxiliary USt codes associated with groups of correctable errors on
size-$t$ clusters.  For a generic non-additive code, this reduces the
number of error-correcting measurements approximately $3^{t}$ times,
compared to exhaustive screening of all correctable errors of weight
$t$ and smaller.

\textsc{Acknowledgment\/}. This research was supported in part by the
NSF grant No.\ 0622242. We are grateful to Bei Zeng for the detailed
explanation of the CWS code construction and to Markus Grassl for 
the important comments.

%\bibstyle{apsrevX}

%\bibliographystyle{apsrev}
%\bibliographystyle{aip} %%% use aip.bst for final version 
%\bibliography{qc_all}

\end{document}